\documentclass[aps,prb,showpacs,reprint,superscriptaddress]{revtex4-1}

\usepackage{graphicx}
\usepackage{color}
\usepackage{amsmath}
\usepackage{bbm}
\usepackage[caption=false]{subfig}

\begin{document}

\title{Impact of the spin-orbit interaction on the phase diagram and anisotropy of the in-plane critical magnetic field in superconducting LaAlO$_3$/SrTiO$_3$ interface}

\author{P. W\'ojcik}
\email{pawel.wojcik@fis.agh.edu.pl}
\affiliation{AGH University of Science and Technology, Faculty of
Physics and Applied Computer Science, 
Al. Mickiewicza 30, 30-059 Krakow, Poland}

\author{M. P. Nowak}
\author{M. Zegrodnik}
\affiliation{AGH University of Science and Technology, Academic Centre for Materials and Nanotechnology, Al. Mickiewicza 30, 30-059 Krakow, Poland}

\date{\today}

\begin{abstract}
The two-dimensional electron gas at the interface between LaAlO$_3$ and SrTiO$_3$ (LAO/STO) exhibits gate tunable superconductivity with a characteristic  
dome-like shape of the critical temperature ($T_c$) in the phase diagram. As shown recently [Phys. Rev. B 102, 085420 (2020)], such an effect can be explained as a consequence of the extended $s-$wave symmetry of the gap within an intersite real space pairing scenario, leading to a good agreement between the experiment and theory. In this work, we turn to a detailed analysis of the influence of spin-orbit coupling on 
the LAO/STO phase diagram by considering the atomic and the Rashba components. In particular, we analyze the optimal carrier concentration for which the maximal $T_c$ is reached relative to the Lifshitz transition point. We find that the a misalignment between the two can be significantly enhanced by the spin-orbit splitting of the bands, combined with the fact that superconductivity sets in when the Fermi level passes the anticrossing induced by the spin-orbital hybridization. In the presence of the external in-plane magnetic field, our calculations show four-fold anisotropy with the paramagnetic limit largely exceeded for $B_{||}$ directed along the high symmetry points [01] and [01].  The obtained electron concentration dependence of $B_{c||}$ reproduces the characteristic dome-like shape reported in experiments and the estimated value of $B_{c||}$ corresponds to that measured experimentally. 
\end{abstract}

\maketitle

\section{Introduction}
In  recent years, the two-dimensional electron gas (2DEG) at the interface between LaAlO$_3$ (LAO) and SrTiO$_3$ (STO) has 
attracted growing interest as a natural platform for studying the interplay between 
superconductivity~\cite{joshua2012universal,maniv2015strong,Biscaras,Monteiro2019,Monteiro2017}, 
magnetism~\cite{Karen2012,Li2011,Brinkman2007,Dikin2011,Bert2011}, spin-orbit 
interaction~\cite{Diez2015,Rout2017,Caviglia2010,Shalom2010,Yin2020,Singh2017,Hurand2015} as well as 
ferroelectricity\cite{Gastiasoro2020,Kanasugi2018,Kanasugi2019,Honig2013}. The 2DEG between LAO and STO results from the polar discontinuity and the 
corresponding charge transfer to the interface that confines electrons within a few TiO$_2$ monolayers\cite{Popovic2008,Pavlenko2012,Pentcheva2006}. 
The low energy electronic structure of the system comes from the $t_{2g}=\{d_{xy},d_{yz},d_{xz}\}$ orbitals of the Ti ions which form a square lattice structure\cite{Delugas2011,Zhong2013,Khalsa2013,joshua2012universal,smink2017}. The spin degeneracy of the resulting three bands is lifted by the inversion symmetry breaking leading to Rashba and atomic spin-orbit coupling (SOC) which affect both the transport and superconducting properties of the 
system\cite{Rout2017,Singh2017,Caviglia2010,Shalom2010,Yin2020,Hurand2015}.

Although superconductivity at the LAO/STO interface has been discovered already some time ago\cite{Reyren2007, joshua2012universal}, it is still insufficiently explored and remains under a wide debate concentrated around the determination of the pairing mechanism\cite{Gorkov2016, Ruhman2016, Breitschaft, Valentinis2017, Klimin2014, Appel1969, Stashans2003, Edge2015, Gamboa2018, Gabay2017, Pekker2020, Sumita2020} and the explanation of unconventional superconducting properties seen in experiments. Among them, one of the most interesting is the dome-like shape of the superconducting transition temperature $T_c$ as a function of the gate potential\cite{Rout2017,joshua2012universal, Caviglia2010, maniv2015strong}, with $T_c$ maxima appearing near the  Lifshitz transition (LT) when an additional band crosses the Fermi level\cite{joshua2012universal, smink2017, 
Singh2017}. 

Up to date, a few  theoretical proposals have been made to explain the characteristic shape of $T_c$. Scenarios that are being considered emphasise the role of electron-electron correlations\cite{Monteiro2019,maniv2015strong}, the strong pair-breaking effect resulting from an interband repulsive interaction\cite{Trevisan2018, singh2019gap} as well as the intersubband pairing in multiband models\cite{Wojcik2020, Mohanta2015}. 
Most of them, however, do not reproduce the experimental data with sufficient agreement. Good correspondence between theory and experiment has been obtained within the model based on the interband scattering~\cite{Singh2017} with the $s_{\pm}$ pairing analogues to that applied to the Fe-based superconductors\cite{Hirshfeld2011}. As shown, the scattering processes between bands corresponding to opposite signs of the gap, lead to Cooper pair-breaking and suppression the critical temperature\cite{Vavilov2011} after the Lifshitz transition is reached\cite{Vavilov2011}. In this scenario, the optimal electron concentration ($n_{\mathrm{opt}}$), for which $T_c$ maximum appears, is located at the LT point. Some experimental analysis confirm such correspondence between LT and $n_{\mathrm{opt}}$\cite{joshua2012universal,singh2019gap}. However, other reports point to a situation in which LT appears before the maximum of $T_C$\cite{Biscaras,maniv2015strong,Caviglia2010,Monteiro2019}. In particular, in Ref. \onlinecite{Biscaras} it is reported that the multiband behavior induced by the LT sets in together with the creation of the SC state. The latter scenario is consistent with our recent theoretical proposal which leads to a good agreement between theory and experiment\cite{Zegrodnik2020}. Within our approach the appearance of the dome-like behavior of $T_c$ as a function of the carrier concentration is a consequence of the extended $s-$wave symmetry of the gap and the topology of the Fermi surface. Based on the Gutzwiller method~\cite{Zegrodnik2020}, we have also found that for the (001)-oriented LAO/STO, the electronic correlations do not play a crucial role in the creation of the $T_c-$dome.

Here we further explore our hypothesis based on the extended $s-$wave symmetry of the gap in the LAO/STO interface. In particular, we analyze the influence of SOC on the phase diagram in the context of the LT and $n_{\mathrm{opt}}$ relative position. Moreover, the interplay between superconductivity and SOC is investigated here, in the presence of the external magnetic field. The latter becomes important in relation to the recent experimental reports\cite{Caviglia2010,Rout2017,Singh2017}. In particular, the SOC leads to an increase of the in-plane critical field up to 2-3~T which is several times greater that the paramagnetic limit\cite{Rout2017,BenShalom2010}. Furthermore, the presence of both SOC and magnetic field may lead to modification of the gap symmetry, which can be crucial for inducing topological features in the superconducting phase\cite{Daido2016}.

First, we show, that due to the appearance of the SOC the system encounters two Lifshitz transitions, as the carrier concentration is increased. The superconducting phase is created in close proximity to the first one, while the $T_c$ maximum appears slightly above the second. We discuss this result in view of the available experimental data. Next, we turn to the analysis of the critical in-plane magnetic field where we find a four-fold anisotropy of $B_{c||}$ with respect to the magnetic field orientation. Interestingly, for a high symmetry crystallographic directions, the estimated value of $B_{c||}$ corresponds to the one measured in the experiment~\cite{Rout2017} ($\sim 4$~T) and is several times higher than the paramagnetic limit. Also, in the presence of the external magnetic field, a spin-triplet $p_x/p_y$ component of the superconducting gap appears due to SOC. 

The paper is organized as follows. In Sec.~\ref{sec2} a theoretical model is provided and the spin-orbit components at the LAO/STO interface are introduced. The influence of SOC on the phase diagram and superconducting state is discussed in sec.~\ref{sec3a}. In Sec.~\ref{sec3b} we analyze the anisotropy of the critical magnetic field resulting from the presence of SOC. Sec.~\ref{sec4} contains the summary.

\section{Theoretical model}  
\label{sec2}
The effective band structure of the (001)-oriented LAO/STO interface is formed by the Ti $t_{2g}$ orbitals confined in a narrow quantum well, which is created near the interface as a result of the polarization discontinuity\cite{Popovic2008,Pavlenko2012,Pentcheva2006}. Here, we supplement the corresponding three-band Hamiltonian ($\hat{H}_{TBA}$)\cite{Diez2015,Khalsa2013} with the coupling to the 
magnetic field ($\hat{H}_{B}$), the Coulomb repulsion ($\hat{H}_{U}$), and the real-space pairing ($\hat{H}_{SC}$). As a result, we obtain 
\begin{equation}
    \hat{H}=\hat{H}_{TBA}+\hat{H}_B+\hat{H}_U+\hat{H}_{SC},
    \label{eq:Hamiltonian_general}
\end{equation}
with the single particle part, $\hat{H}_{TBA}$, which takes the form
\begin{equation}
 \hat{H}_{TBA}=\sum _{\mathbf{k}l\sigma} \hat{c}^{\dagger}_{\mathbf{k},l,\sigma} \left ( \hat{H}_0 + \hat{H}_{SO} + \hat{H}_{RSO}\right ) 
\hat{c}_{\mathbf{k}l\sigma},
\label{h:tba}
\end{equation}
where $\hat{c}^{\dagger}_{\mathbf{k}l\sigma}(\hat{c}_{\mathbf{k}l\sigma})$ creates (annihilates) electrons with spin $\sigma$, momentum 
$\mathbf{k}$, and the band index $l=xy,xz,yz$. The latter corresponds to the three orbitals $d_{xy}, d_{xz}, d_{yz}$ of the Ti atoms placed on the square lattice. \\
$\hat{H}_0$ term describes the three bands of the system
\begin{equation}
\hat{H}_{0}=
\left(
\begin{array}{ccc}
 \epsilon^{xy}_{\mathbf{k}} & 0 & 0\\
 0 & \epsilon^{xz}_{\mathbf{k}} &  \epsilon^h_{\mathbf{k}} \\
 0 & \epsilon^h_{\mathbf{k}}  &  \epsilon^{yz}_{\mathbf{k}}
\end{array} \right) \otimes \hat {\sigma} _0\;,
\end{equation}
with the dispersion relations 
\begin{equation}
\begin{split}
    \epsilon^{xy}_{\mathbf{k}}&=4t_l-\Delta_E-2t_l\cos{k_x}-2t_l\cos{k_y},\\
    \epsilon^{xz}_{\mathbf{k}}&=2t_l+2t_h-2t_l\cos{k_x}-2t_h\cos{k_y},\\
    \epsilon^{yz}_{\mathbf{k}}&=2t_l+2t_h-2t_h\cos{k_x}-2t_l\cos{k_y},
\end{split}
\label{eq:H0}
\end{equation}
and the hybridization term 
\begin{equation}
\epsilon^h_{\mathbf{k}}=2t_d\sin{k_x}\sin{k_y}. 
\end{equation}
The tight-binding parameters are reported in Ref. \onlinecite{maniv2015strong} and 
take the values: $t_l=875\;$meV, $t_h=40\;$meV, $t_d=40\;$meV, $\Delta_E=47\;$meV.

The remaining components of Eq. (\ref{h:tba}) are related to the atomic and Rashba SOC, respectively.
The former appears as an effect of the crystal field splitting of the atomic orbitals and is given by\cite{Khalsa2013}
\begin{equation}
\hat{H}_{SO}= \frac{\Delta_{SO}}{3}
\left(
\begin{array}{ccc}
0 & i \hat{\sigma _x} & -i \hat{\sigma _y}\\
-i \hat{\sigma _x} & 0 & i \hat{\sigma _z} \\
i \hat{\sigma _y} & -i \hat{\sigma _z} & 0
\end{array} \right) \;,
\label{eq:hso}
\end{equation}
where $\Delta _{SO}$ determines the atomic-like spin-orbit energy and $\mathbf{\sigma} = (\sigma _{x},\sigma_{y},\sigma_{z})$ are the Pauli matrices. 

The Rashba spin-orbit term $\hat{H}_{RSO}$ results from the intrinsic electric field at the interface which breaks the inversion symmetry and takes the form
\begin{equation}
\hat{H}_{RSO}= \Delta_{RSO}
\left(
\begin{array}{ccc}
0 & i \sin{k_y} & i \sin{k_x}\\
-i \sin{k_y} & 0 & 0 \\
-i \sin{k_x} & 0 & 0
\end{array} \right) \otimes \hat {\sigma} _0\;,
\label{eq:rso}
\end{equation}
with $\Delta _{RSO}$ controlling the strength of the interaction. \\

In Eq.~(\ref{eq:Hamiltonian_general}), the coupling of the external magnetic field to the spin and orbital momentum of electrons is included by $\hat{H}_B=\mu_B(\mathbf{L}+g\mathbf{S})\cdot \mathbf{B}/\hbar$ with $g=5$, $\mathbf{S}=\hbar \pmb{\sigma}/2$, where
\begin{equation}
\begin{split}
 L_x&=\left ( 
 \begin{array}{ccc}
  0 & i & 0 \\
  -i & 0 & 0 \\
  0 & 0 & 0 
 \end{array}
 \right ), 
 L_y=\left ( 
 \begin{array}{ccc}
  0 & 0 & -i \\
  0 & 0 & 0 \\
  i & 0 & 0 
 \end{array}
 \right ), \\
 L_z&=\left ( 
 \begin{array}{ccc}
  0 & 0 & 0 \\
  0 & 0 & i \\
  0 & -i & 0 
 \end{array}
 \right ).
 \end{split}
\end{equation}
The Coulomb repulsion term $\hat{H}_U$ has the following form
\begin{equation}
 \hat{H}_{U}=U\sum_{il}\hat{n}_{il\uparrow}\hat{n}_{il\downarrow}+V\sideset{}{'}\sum_{ill'}\hat{n}_{il}\hat{n}_{il'},
 \label{eq:Hamiltonian_U}
\end{equation}
where $U$ and $V$ are the intra- and inter-orbital Coulomb repulsion integrals while the primmed summation is restricted to $l\neq l'$. For simplicity, we take $U=V\equiv2\;$eV, which corresponds to the value calculated in Ref. \onlinecite{Breitschaft}.

Finally, in our model, the superconducting state is enabled by a real-space intersite intraorbital pairing between the nearest neighboring sites as well as the interorbital pair hopping term
\begin{equation}
\begin{split} 
\hat{H}_{SC}=&-J\sum_{\langle ij \rangle l\sigma}\hat{c}^{\dagger}_{il\sigma}\hat{c}^{\dagger}_{jl\overline{\sigma}}\hat{c}_{il\overline{\sigma}}\hat{c}_{
jl\sigma } 
-J^{\prime}\sideset{}{'}\sum_{\langle ij \rangle ll'}\hat{c}^{\dagger}_{il\sigma}\hat{c}^{\dagger}_{jl\overline{\sigma}}\hat{c}_{il'\overline{\sigma}}\hat{c}_{
jl'\sigma },
 \end{split}
 \label{eq:Hamiltonian_pairing}
\end{equation}
where the summations run over the nearest-neighbors only, $\overline{\sigma}$ is the spin opposite to $\sigma$, and the pairing strength is determined by the parameters $J$ and $J^{\prime}$, respectively. In the calculations, we assume that the interorbital pair hopping energy, $J^{\prime}$, 
is one order of magnitude smaller than the intraorbital coupling constant, $J$.

Within the standard mean field approach, Hamiltonian (\ref{eq:Hamiltonian_general}) can be transformed to
\begin{equation}
\begin{split}
\hat{H}=&\frac{1}{2}\sum_{\mathbf{k}}\mathbf{\hat{f}}^{\dagger}_{\mathbf{k}}\mathbf{\hat{H}}_{\mathbf{k}}\mathbf{\hat{f}}_{\mathbf{k}}+	
\frac{1}{2}\sum_{\mathbf{k}ls=\pm 1} (\xi^l_{\mathbf{k}} + \frac{1}{2} g \mu_B s B_z) \\
&-N\sum_{l}\bigg(U n_{l\uparrow}n_{l\downarrow}+V\sum_{ l'(l'\neq l)}n_{l}n_{l'}\bigg) \\
&+\sum _{\langle ij \rangle l\sigma} J  \langle \hat{c}_{il\sigma}\hat{c}_{jl\overline{\sigma}}\rangle ^2 \\
& + \sum_{\langle ij \rangle ll'\sigma} J'\langle \hat{c}_{il\sigma}\hat{c}_{jl \overline{\sigma}}\rangle \langle \hat{c}_{il'\sigma}\hat{c}_{jl' 
\overline{\sigma}'} \rangle,
\end{split}
\label{eq:ham_HF}
\end{equation}
where $N$ is the number of atomic sites in our system, 
$n_l=\langle\hat{n}_{il\uparrow}\rangle+\langle\hat{n}_{il\downarrow}\rangle$, while the vector $\mathbf{\hat{f}}_{\mathbf{k}}$ is the twelve-component composite operator
\begin{equation}
\begin{split}
 \mathbf{\hat{f}}^{\dagger}_{\mathbf{k}}\equiv (&\hat{c}^{\dagger}_{\mathbf{k},xy\uparrow},\hat{c}^{\dagger}_{\mathbf{k},xy\downarrow}, 
 \hat{c}^{\dagger}_{\mathbf{k},xz\uparrow},\hat{c}^{\dagger}_{\mathbf{k},xz\downarrow},
 \hat{c}^{\dagger}_{\mathbf{k},yz\uparrow},\hat{c}^{\dagger}_{\mathbf{k},yz\downarrow}, \\
&\hat{c}_{-\mathbf{k},xy\uparrow},\hat{c}_{-\mathbf{k},xy\downarrow}, 
\hat{c}_{-\mathbf{k},xz\uparrow}, \hat{c}_{-\mathbf{k},xz\downarrow}, 
\hat{c}_{-\mathbf{k},yz\uparrow}, \hat{c}_{-\mathbf{k},yz\downarrow})\;.
\end{split}
\end{equation}
The full matrix form of $ \mathbf{\hat{H}}_{\mathbf{k}}$ is given by
\begin{equation}
\begin{split}
 \mathbf{\hat{H}}_{\mathbf{k}}=\sigma _z \otimes H_0'+ \sigma_0 \otimes& H_{RSO}+ \sigma_z \otimes H_{SO}\\
 +& \sigma _z \otimes H_B + \sigma _x \otimes 
\Gamma_{6\times 6},
\end{split}
\label{eq:matrix_H}
\end{equation}
where the prime of $H_0$ in Eq.~(\ref{eq:matrix_H}) indicates that in the diagonal elements $\epsilon^l_{\mathbf{k}}$, Eqs. (\ref{eq:H0}), we include the chemical potential term as well as the effective shift of the atomic energy, which originates from the Hartree-Fock approximation of the Coulomb interaction terms. Namely, in $H_0'$ the diagonal elements of Eq.~(\ref{eq:matrix_H}) are replaced by
\begin{equation}
    \xi^ l_{\mathbf{k}}=\epsilon^l_{\mathbf{k}}+U\frac{n_{il}}{2}+V\sum_{l'(l'\neq l)}n_{l'}-\mu.
    \label{eq:diagonal_dissp}
\end{equation}
The part of $\mathbf{\hat{H}}_{\mathbf{k}}$ which is responsible for superconductivity takes the form
\begin{equation}
\Gamma_{6\times6}=\left(\begin{array}{cccccc}
 0 & \Gamma _{xy}^{\uparrow \downarrow} & 0 & 0 & 0 & 0\\
\Gamma _{xy}^{\downarrow \uparrow} & 0 & 0 & 0 & 0 & 0\\
0 & 0 & 0 & \Gamma _{xz}^{\uparrow \downarrow} & 0 & 0 \\
0 & 0 & \Gamma _{xz}^{\downarrow \uparrow} & 0 & 0 & 0 \\
0 & 0 & 0 & 0 & 0 & \Gamma _{yz}^{\uparrow \downarrow}\\
0 & 0 & 0 & 0 & \Gamma _{yz}^{\downarrow \uparrow} & 0 \\
\end{array} \right)\;,
\label{eq:matrix}
\end{equation}
where
\begin{eqnarray}
 \Gamma_{l}^{\sigma \overline{\sigma}}= \frac{1}{2} \sum _{\langle j(i)\rangle} \Delta_{ijl}^{\sigma \overline{\sigma}}\; e^{i\mathbf{k}\cdot \mathbf{R}_{ij}},
 \label{eq:gamma}
\end{eqnarray}
and $\mathbf{R}_{ij}=\mathbf{R}_{i}-\mathbf{R}_{j}$. The summation above runs over the four nearest-neighboring atomic sites surrounding the $i$-th atomic site. Value of $\Gamma_{l}^{\sigma \overline{\sigma}}$ does not depend on the position of $i$-th atomic site since we are considering a spatially homogeneous system ($\Gamma_{il}^{\sigma \overline{\sigma}}\equiv \Gamma_{l}^{\sigma \overline{\sigma}}$). The explicit definition of the pairing amplitudes $\Delta_{ijl}^{\sigma \overline{\sigma}}$ is given by
\begin{equation}
\begin{split}
   \Delta_{ijl}^{\sigma \overline{\sigma}}=-J\langle\hat{c}_{il\sigma}\hat{c}_{jl 
\overline{\sigma}}\rangle-J^{\prime}\sum_{l'(l'\neq l)}\langle\hat{c}_{il'\sigma}\hat{c}_{jl'\overline{\sigma}}\rangle.
   \label{eq:Delta_real_space}
\end{split}
\end{equation}

Even though, in the absence of the external magnetic field, the extended $s$-wave pairing amplitude appears as stable\cite{Zegrodnik2020}, the inclusion of both SOC and the external magnetic field may have a significant effect on the gap symmetry. Therefore, here we consider the $s$, $p$, and $d$ symmetry resolved pairing amplitudes which can be determined from Eq. (\ref{eq:gamma}) by using the values of $\Delta_{ijl}$. They have the form
\begin{equation}
    \Delta^{\sigma\bar{\sigma}|s,p,d}_{l}=\frac{1}{4}\sum_{\langle j(i) \rangle}\gamma^{s,p,d}_{ij}\Delta_{ijl}^{\sigma \overline{\sigma}},\quad     
    \label{eq:Delta_amplitudes}
\end{equation}
where the real-space symmetry factors are 
\begin{equation}
\begin{split}
\gamma^{s}_{ij}&=(\delta_{\mathbf{R}_{ij}-\hat{x}}+\delta_{\mathbf{R}_{ij}+\hat{x}}+\delta_{\mathbf{R}_{ij}-\hat{y}}+\delta_{\mathbf{R}_{ij}+\hat{y}}),\\
\gamma^d_{ij}&=(\delta_{\mathbf{R}_{ij}-\hat{x}}+\delta_{\mathbf{R}_{ij}+\hat{x}}-\delta_{\mathbf{R}_{ij}-\hat{y}}-\delta_{\mathbf{R}_{ij}+\hat{y}}),\\
\gamma^{p_x}_{ij}&=(\delta_{\mathbf{R}_{ij}-\hat{x}}-\delta_{\mathbf{R}_{ij}+\hat{x}}),\\
\gamma^{p_y}_{ij}&=(\delta_{\mathbf{R}_{ij}-\hat{y}}-\delta_{\mathbf{R}_{ij}+\hat{y}}),\\
\end{split}
\label{eq:symmetries}
\end{equation}
with $\delta_{\mathbf{v}}$ being the appropriate Kronecker delta, giving $\delta_{\mathbf{v}}=1$ only when $\mathbf{v}=(0,0)$ and $\hat{x} (\hat{y})$ is the unit vector along $x(y)$ axis. It should be noted, that the spin-singlet pairing ($\Delta^{\uparrow\downarrow}=-\Delta^{\downarrow\uparrow}$) is compatible with the $s-$ and $d-$wave gap symmetries, while the spin-triplet pairing ($\Delta^{\uparrow\downarrow}=\Delta^{\downarrow\uparrow}$) is allowed for the $p-$wave gap symmetry. In general, in the considered model, a multicomponent gap structure can be realized with a mixture of $s$, $p$, and $d$ symmetries as well as both singlet and triplet contributions.

The pairing energies $\Gamma^{\sigma \overline{\sigma}}_l$ are determined by solving the set of self-consistent equations (\ref{eq:matrix_H}) and  (\ref{eq:Delta_real_space}). The numerical diagonalization of (\ref{eq:matrix_H}) leads to the quasi-particle energies 
$\lambda_{\mathbf{k}l\sigma}$ which then are used to determine the free energy functional  in  a  standard statistical
manner
\begin{eqnarray}
 F&=&-\frac{1}{2}k_BT \sum _{\mathbf{k} l\sigma} \ln \left [ 1 + \exp \left ( \frac{\lambda_{\mathbf{k}l\sigma}}{k_BT} \right ) \right ]  \nonumber \\
 &+& \frac{1}{2}\sum _{\mathbf{k}ls=\pm 1} \left [ \xi^ l_{\mathbf{-k}} + \frac{1}{2}g\mu_B s B_z - \lambda_{\mathbf{k}l\downarrow}  \right 
]  \nonumber \\
 &-& \frac{1}{2} J \sum _{il\sigma} |\langle\hat{c}_{il\sigma}\hat{c}_{i+\hat{x}(\hat{y})l \overline{\sigma}}\rangle|^2 -  
 \frac{1}{2} J^\prime \sum _{i\sigma l\ne l'}  |\langle\hat{c}_{il\sigma}\hat{c}_{i+\hat{x}(\hat{y})l' \overline{\sigma}}\rangle|^2 \nonumber \\
 &-& U \sum _l  n_{l\uparrow} n_{l\downarrow} - \frac{1}{2} V \sum _{l\ne l'}  n_{l} n_{l'} + \sum _l \mu n_l.
\end{eqnarray}
The superconducting state is thermodynamically stable when its free energy is lower than the energy of a normal state. 

\section{Results}
\label{sec3}
Here we discuss the impact of the atomic and Rashba SOC on the superconducting properties of 2DEG at the LAO/STO interface. Then, we analyze an anisotropy of the critical magnetic field induced by the spin-orbital interaction. Although in experiments SOC can change with the gate potential\cite{Yin2020, BenShalom2010, Caviglia2010}, for the sake of simplicity, we assume that it is constant and does not depend on the electron concentration.

\subsection{Influence of SOC on the phase diagram structure}
\label{sec3a}

\begin{figure}
 \centering
 \includegraphics[width=0.45\textwidth]{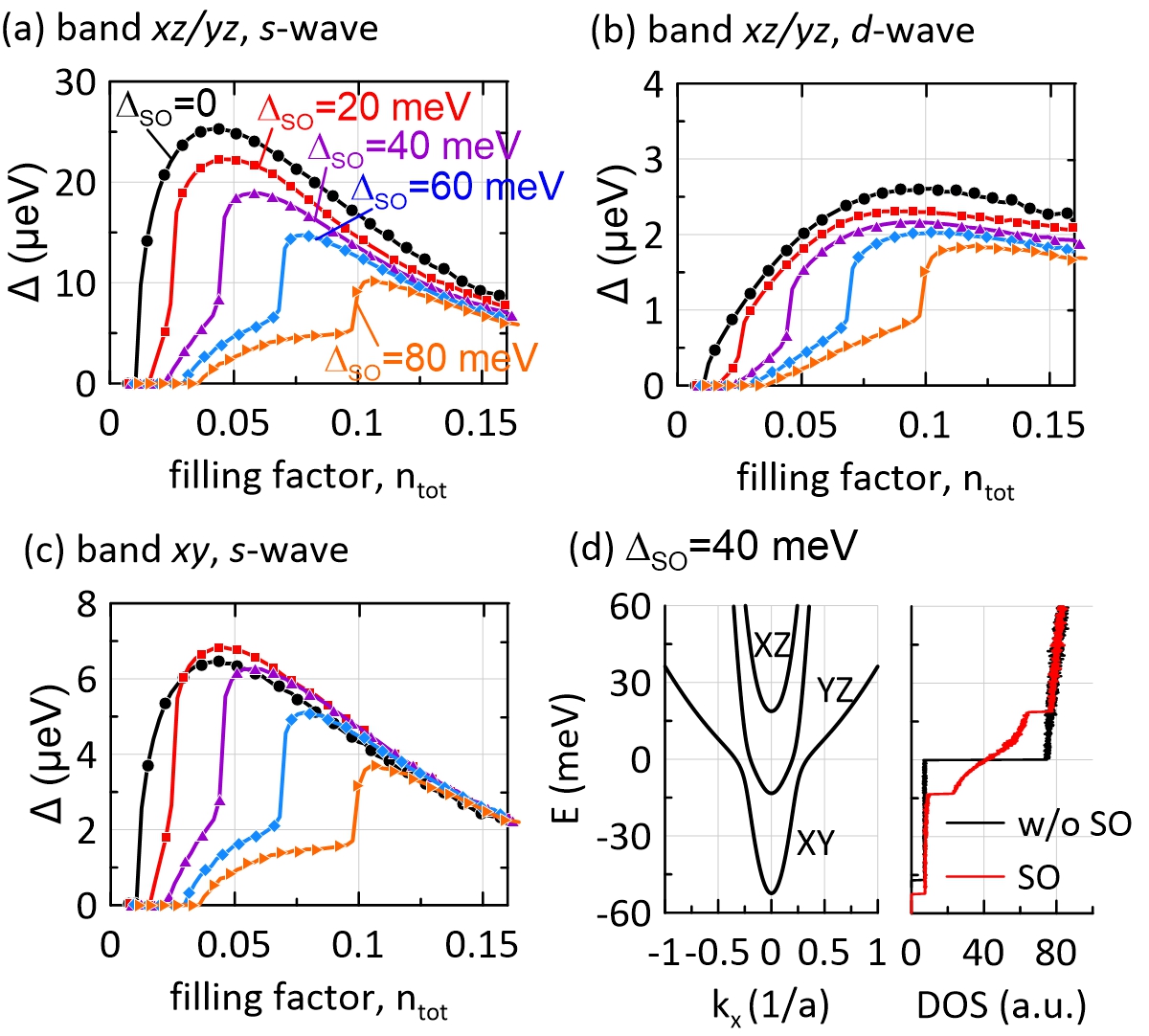}
 \caption{The extended $s-$ and $d-$ wave pairing amplitudes of the two $xz/yz$ bands (a,b) and the $xy$ band (c) [see Eqs. (\ref{eq:Delta_real_space}), (\ref{eq:Delta_amplitudes})] as a function of the band filling $n_{tot}$ for different atomic SOC strengths $\Delta_{SO}$. (d) Band structure of LAO/STO interface at $k_y=0$, for $\Delta_{SO}=40$~meV (left panel) together with the density of states (DOS) (right panel). For comparison, DOS calculated without SOC is plotted by the black line.}
\label{fig1}
\end{figure}

First, we focus on the sole effect of the atomic SOC by assuming $\Delta_{RSO}=0$. We set the superconducting coupling constant $J= 0.165$~eV  ($J'=0.0165$~eV) so that the model with $\Delta_{SO}=0$ reproduces the maximal critical temperature $T_c \approx 0.35$ K measured in the experiment\cite{joshua2012universal}.
Results at $T=0$ for different values of $\Delta_{SO}$ are shown in Fig.~\ref{fig1}(a-c). To better highlight all the effects driven by the atomic SOC, we have carried out our calculations for the $\Delta_{SO}$ parameter up to a value of $80$~meV, which should be considered as relatively high when it comes to the LAO/STO system\cite{Yin2020, BenShalom2010, Caviglia2010}. Note, however, that even stronger spin-orbit splitting energy ($0.5$ ~eV) at the oxide interfaces has been recently reported \cite{KTO} for (111) KTaO$_3$/SrTiO$_3$.

As shown in Fig.~\ref{fig1}, the pairing amplitudes for $xz$ and $yz$ are equal due their hybridization induced by the interorbital mixing term, 
$\epsilon ^h_{\mathbf{k}}$, and SOC. Regardless of $\Delta_{SO}$, the extended $s-$wave pairing  in the $xz/yz$ band constitutes the dominant 
contribution to the superconducting phase. It reproduces the experimentally measured dome-like shape of the critical temperature~\cite{joshua2012universal}. 
Although in the experiment the superconducting dome is measured as a function of the gate potential, $V_g$,  the increase of $V_g$ is equivalent to 
adding electrons to the system and changes the filling factor, $n_{tot}=\sum _l n_l$, used in Fig.~\ref{fig1}.
The $d-$wave component [Fig.~\ref{fig1}(b)] for the bands $xz/yz$ is one order of magnitude smaller than $\Delta _{xz/yz}^s$ and almost vanishes for 
the underlying $xy$ band (not displayed here). Note that, regardless of the type and magnitude of the SOC, for $B=0$, we find $\Delta^{p_x}_l=\Delta^{p_y}_l\equiv 0$.

As implied by Eq.~(\ref{eq:hso}) the atomic SOC mixes spin-opposite bands $xy$ and $xz/yz$. Furthermore, the bands $xz$ and $yz$ are hybridized via the Zeeman-like term proportional to $\sigma_z$. 
As a result, the spin and band indexes are no longer good quantum numbers, and the electronic state at a certain momentum $\mathbf{k}$ is a mixture of the original orbital states with the spin quantum number undefined. To distinguish between the original orbitals $xy, xz, yz$ and the so called helical states created after the inclusion of SOC, the latter will be labeled by $XY, XZ, YZ$.
As shown in Fig.~\ref{fig1}(d), the bands $XZ$ and $YZ$ are split in energy by $\Delta_{SO}$ and an anticrossing is created between the two lowest helical bands.

In Fig. \ref{fig1} we show that the SOC-induced change of the band structure entails the deformation of the superconducting dome with an appearance of a characteristic slow increase of the superconducting amplitudes
below the principal maximum. After the inclusion of SOC, two Lifshitz transitions appear -- LT1 and LT2 [cf. Fig. \ref{fig1}(d)]  corresponding to the two subsequent helical bands $YZ, XZ$ crossing the Fermi level. We found that the rapid increase of the superconducting gap and the corresponding $T_c$ maximum are associated with the second transition (LT2), which is moved towards higher energies (band filling) as we increase $\Delta_{SO}$ (see the discussion below). In addition, the creation of the helical bands makes the increase of the density of states (DOS) more continuous with respect to the $\Delta_{SO}=0$ case (cf. Fig. \ref{fig1}d). One can find a direct correspondence between the structure of DOS and the characteristic behavior of superconducting amplitudes below the $T_c$ maximum. Note, however, that the fall of the superconducting amplitudes above the $T_c$ maximum is not determined by the structure of DOS, which slightly increases with energy above LT2 [Fig.~\ref{fig1}(d)]. Instead, it results from the extended $s-$wave symmetry of the gap as described in Ref.~\onlinecite{Zegrodnik2020}. 
\begin{figure}
 \centering
 \includegraphics[width=0.45\textwidth]{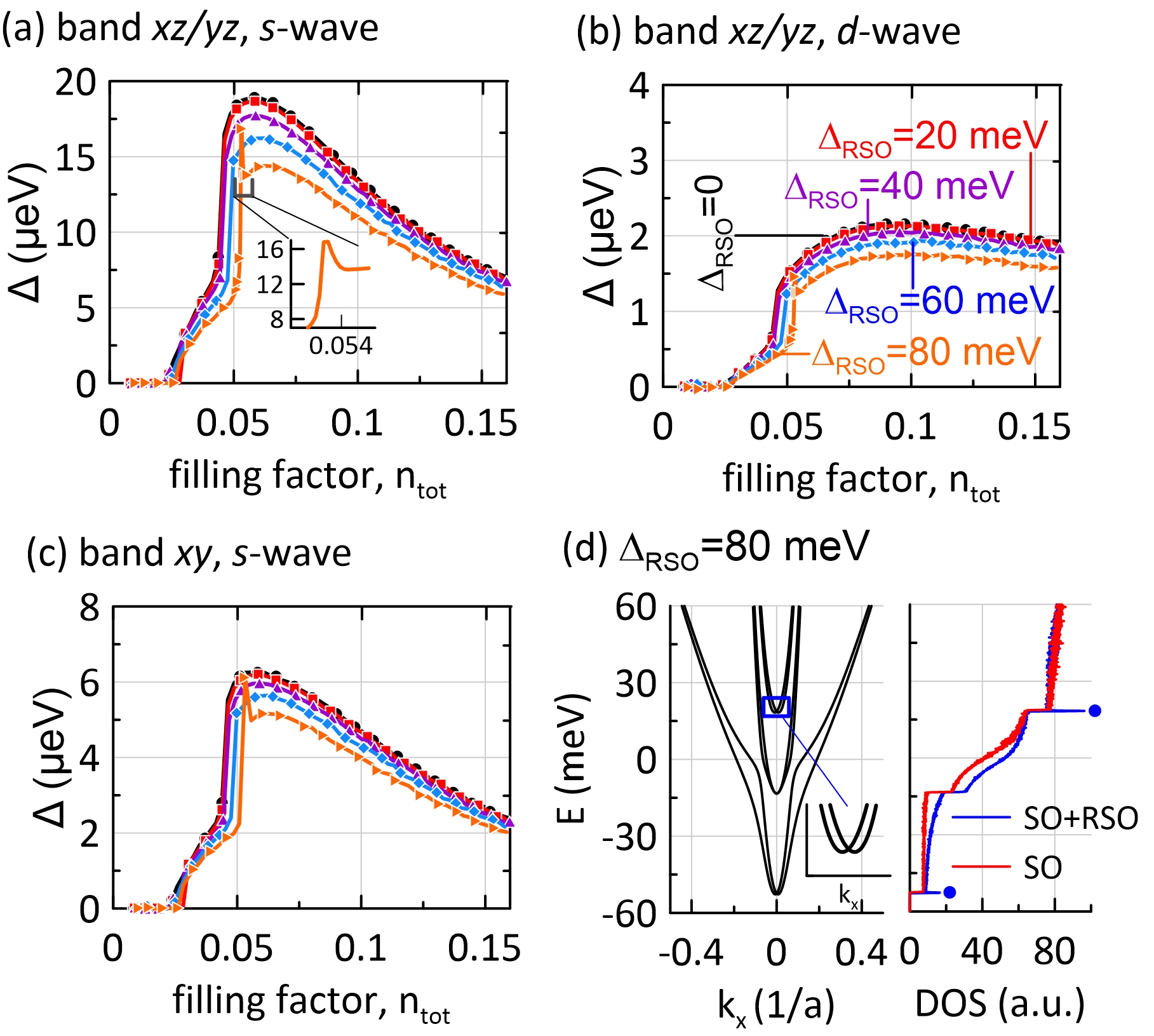}
 \caption{The extended $s-$ and $d-$ wave pairing amplitudes of the two $xz/yz$ bands (a,b) and the $xy$ band (c) [see  Eqs. (\ref{eq:Delta_real_space}), (\ref{eq:Delta_amplitudes})] as a function of the band filling $n_{tot}$, for different Rashba SOC strengths   $\Delta_{RSO}$. The inset in panel (a) presents the enhancement of superconducting amplitude due to the van Hove singularity for $\Delta _{RSO}=80$~meV.
 (d) Band structure of LAO/STO interface at $k_y=0$, for $\Delta_{RSO}=80$~meV (left panel) together with DOS (right panel) with the van Hove singularities marked by blue dots. For comparison, DOS calculated without Rashba SOC is marked by the red line. Results for $\Delta_{SO}=40$~meV.}
\label{fig2}
\end{figure}

Now, let us consider the effect of the Rashba SOC defined by Hamiltonian (\ref{eq:rso}). For his purpose we set $\Delta_{SO}=40$~meV and calculate the superconducting amplitudes for different $\Delta_{RSO}$, as shown in Fig.~\ref{fig2}. 
The impact of the Rashba SO coupling is much weaker than its atomic correspondent -- it only slightly suppresses the superconductivity. As shown in Fig.~\ref{fig2}(d), $\hat{H}_{RSO}$ leads to splitting of the highest helical state parabolas in $k$-space [see the inset in Fig.~\ref{fig2}(d)] and enhances the anticrossing between the lowest helical states. When the bottom of the parabolic band is moved away from $\mathbf{k}=0$, the van Hove singularity appears in DOS\cite{Ast2007}, marked by blue dots in the right panel of Fig.~\ref{fig2}(d). Since larger DOS at the Fermi level leads to stronger superconductivity, the van Hove singularity is manifested as a very narrow peak in the superconducting amplitudes occurring for a strong Rashba SOC -- see the inset in Fig.~\ref{fig2}(a) for $\Delta_{RSO}=80$~meV. Similar effect has been analyzed for an ordinary single-band 2DEG with the Rashba SOC~\cite{Cappelluti2007}.
One should note that the considered effect does not occur for the $XZ$ band since its bottom is located at $\mathbf{k}=0$ and is not affected by the Rashba SOC. Moreover, the singularity corresponding to the band $XY$ is not observed in the superconducting amplitudes since the occupation of the $XY$ band does not induce superconductivity due to insufficient DOS.
\begin{figure}
 \centering
 \includegraphics[width=0.5\textwidth]{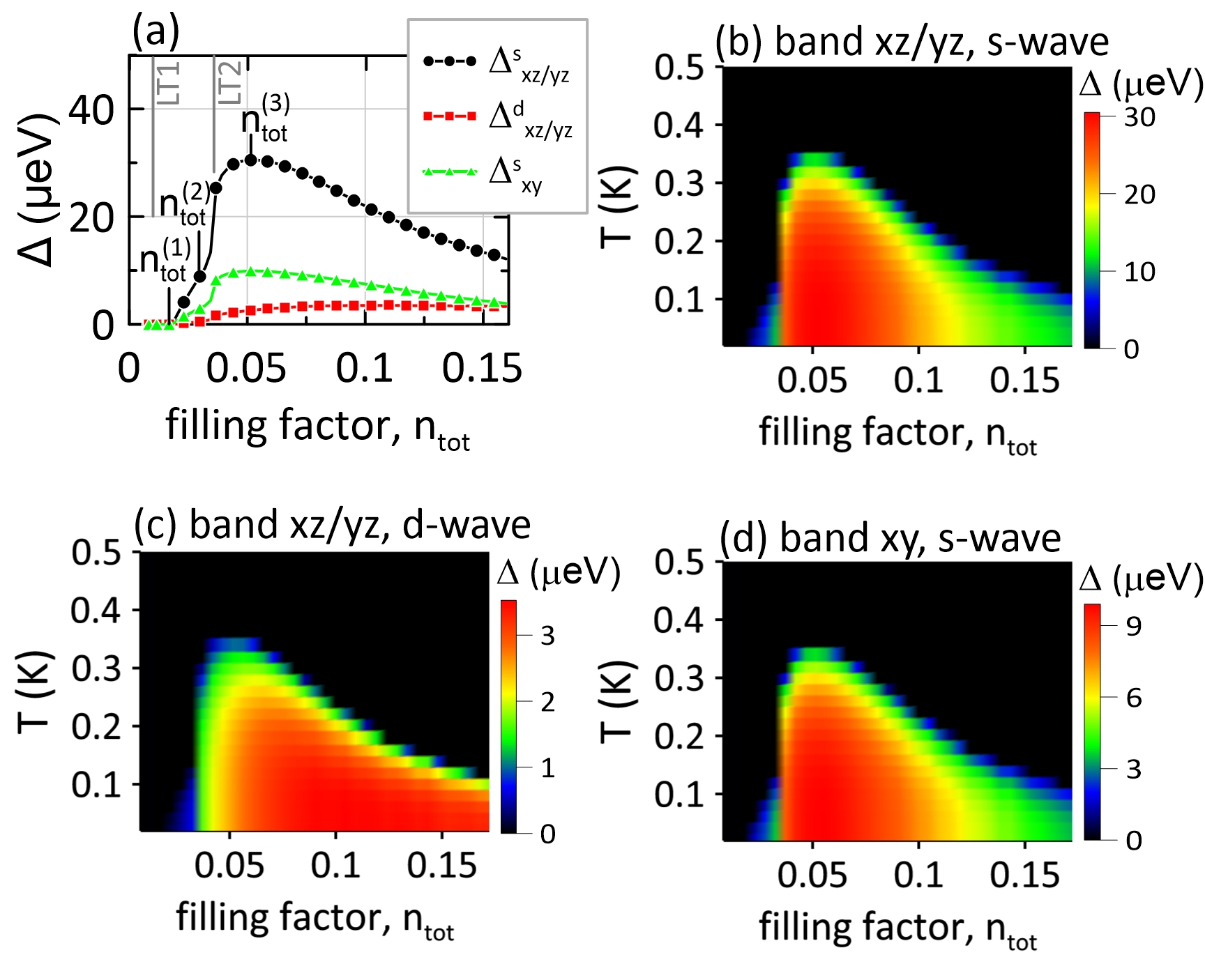}
 \caption{ (a) The zero-temperature extended $s-$wave and $d-$wave pairing amplitudes for all the orbitals of the model as a function of the filling factor, $n_{tot}$. On the top axis two values of the filling factor corresponding to the first and second Lifshitz transitions are marked. Panels (b-d) present the temperature dependencies of the corresponding superconducting amplitudes. Results for $J=0.175$~eV ($J'=0.0175$~eV), $\Delta_{SO}=30$~meV and $\Delta_{RSO}=30$~meV.}
\label{fig3}
\end{figure}

After the analysis of the influence of the two SOC components on the superconducting state, we now discuss the temperature-dependent phase diagrams calculated for realistic values of both SOC parameters $\Delta_{SO}=30$~meV and $\Delta_{RSO}=30$~meV~\cite{Yin2020, BenShalom2010, Caviglia2010}. 
Again, we choose the coupling constants so that the model with SOC reproduces the maximal critical temperature $T_c \approx 0.35$ K measured in the experiment~\cite{joshua2012universal}. The resulting values are $J= 0.175$~eV and $J'=0.0175$~eV. In Fig.~\ref{fig3}(b-d) we show that indeed the dome-like shape of $T_c$ as a function of the filling factor is reproduced in our model and agrees very well with the experiment (compare with Fig.~4 in Ref.~\onlinecite{joshua2012universal}). As can be seen, the superconducting amplitudes at $T=0$~K [panel (a)] differ in magnitude for the three orbitals taken into account in the model. However, the Cooper pair-hopping term with $J'$ connects all of them and guarantees the appearance of a single critical temperature.

The appearance of dome-like shape behavior of $T_c$ in the system without SOC was explained in detail in our recent paper\cite{Zegrodnik2020}, as resulting from the interplay between the topology of the Fermi surface and the gap symmetry. In general, this mechanism does not change in the model with SOC. However, the inclusion of SOC sheds a new light on the physical explanation of the phase diagram, especially when it comes to the relative position of $n_{\mathrm{opt}}$ in relation to the Lifshitz transition. This issue is especially important because it distinguishes our theoretical proposal based on the extended $s-$wave symmetry\cite{Zegrodnik2020} from the one which emphasises the interband scattering with the $s_{\pm}-$pairing\cite{singh2019gap}. As mentioned in the introduction, some experimental reports point to a situation in which the optimal carrier concentration appears exactly at LT\cite{joshua2012universal,singh2019gap}, supporting the interband scattering scenario\cite{Trevisan2018} where the pair breaking should appear after the second band crosses the Fermi level. However, other experimental analysis suggest the location of $n_{\mathrm{opt}}$ above LT\cite{Biscaras,maniv2015strong,Caviglia2010,Monteiro2019} which is consistent with our approach\cite{Zegrodnik2020}. Up to date, there is no common consensus if the maximum of $T_c$ occurs exactly at the Lifshitz point or is located above it. Within the considered model with the SO coupling, the situation related with the latter issue is the following.
\begin{figure}[t]
 \centering
 \includegraphics[width=0.4\textwidth]{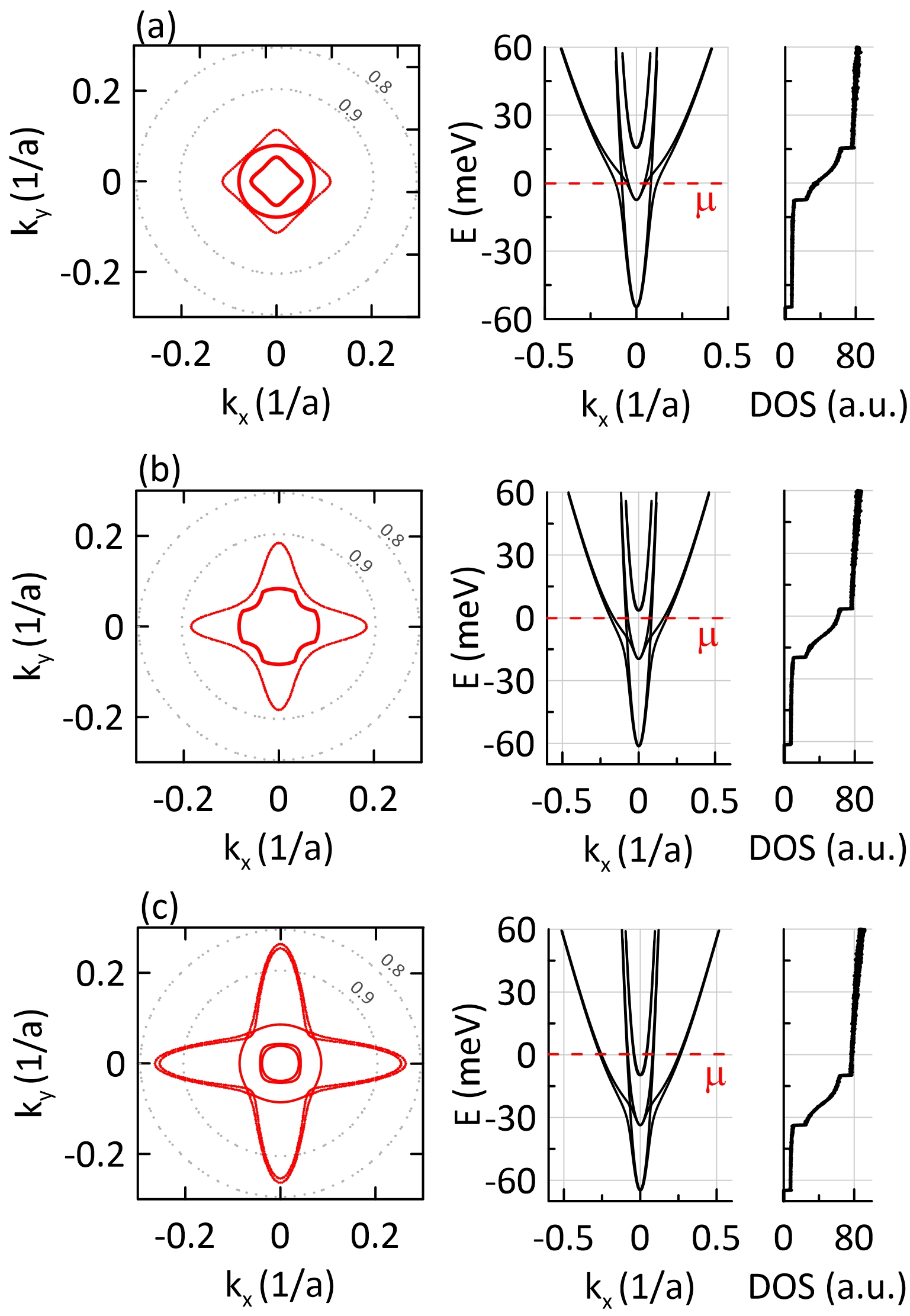}
 \caption{(a-c) Fermi surfaces (left panels), dispersion relations $E(k_x,k_y=0)$ (middle panels) and DOS (right panels) calculated for three values of the filling factor $n_{tot}^{(i)}$ marked in Fig.~\ref{fig3}(a). 
 (d) Band filling components corresponding to the $xy$ and $xz/yz$ bands. The first Lifshitz transition (LT1) is denoted by the dashed vertical line.}
\label{fig4}
\end{figure}

In panel (a) of Fig.~\ref{fig3} we mark three characteristic filling factor values, $n_{tot}^{(i)}$. Namely, (i) just below the onset of superconductivity, (ii) inside the region of moderate increase of the pairing amplitudes which results from the atomic SOC [Fig.~\ref{fig1}], and (iii) at $T_c$ maximum, respectively. 
The Fermi surfaces, together with the Fermi energy position at the dispersion relations $E(k_x,k_y=0)$ and DOS for the chosen values of $n_{tot}^{(i)}$ are presented in Fig.~\ref{fig4}. Even though $n_{tot}^{(1)}$ is already in the region, when the helical state $XZ$ crosses the Fermi level (after the first Lifshitz transition, LT1), the superconductivity is still not induced [Fig.~\ref{fig4}(a)].
Instead, the paired state is created slightly above LT1 when the Fermi energy is placed at the anticrossing. 
At the latter, the band $xz$ with the larger effective mass hybridizes with the $xy$ band, leading to the growth of DOS which becomes sufficient to induce superconductivity. 
For $n^{(2)}_{tot}$ [cf. panel (b)], the Fermi energy $\mu$ is above the anticrossing but still below the second Lifshitz transition (LT2). Although the system exhibits superconductivity, its $T_c$ is very low and far from the maximal value [cf. Fig.~\ref{fig3}]. The rapid growth of $T_c$ occurs at LT2 corresponding to the band $YZ$ and results from the step-like increase of DOS presented in Fig.~\ref{fig4}.

As shown in our recent paper~\cite{Zegrodnik2020}, the experimentally observed misalignment between the Lifshitz transition and the $T_c$ maximum\cite{joshua2012universal,Biscaras} is the intrinsic feature of the model based on the extended $s-$wave gap symmetry. Above, we demonstrated that the distance between both the points can be significantly enhanced by the spin-orbit interaction. 
Due to the latter, the system actually encounters two Lifshitz transitions, as the carrier concentration is increased. The superconducting phase is created slightly above the first one, at the anticrossing,  while the $T_c$ maximum appears slightly above the second.  
An important question would be which of the two transitions are actually measured in particular experiments. In this respect, the crucial issue is related with the accuracy of the experimental determination of $T_c$ maximum and LT. It is because, all those interesting features in the phase diagram (LT1, LT2, optimal concentration, and the SC critical concentration) are relatively close to each other. Nevertheless, it may be the case that the seemingly contradictory sets of experimental data showing two different positions of LT\cite{joshua2012universal,Biscaras} may actually be in agreement in the view of our theoretical results.

\subsection{Critical magnetic field anisotropy}
\begin{figure*}[t]
 \centering
 \includegraphics[width=0.8\textwidth]{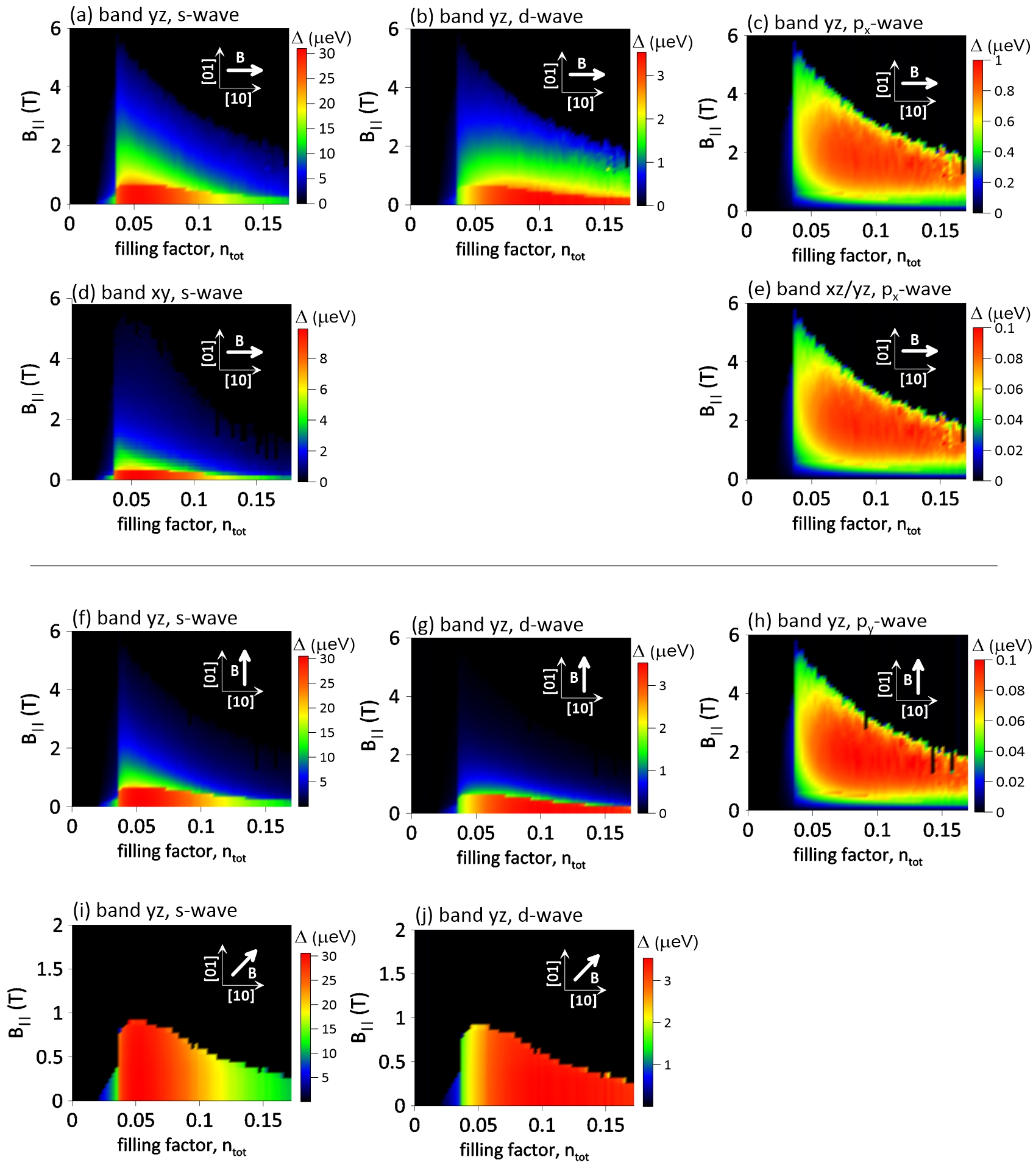}
 \caption{(a-e) The temperature dependencies of the extended $s-$wave, $d-$wave and $p_{x(y)}-$wave pairing amplitudes for the band $yz$ and the lower lying $xy$, as a function of the filling factor $n_{tot}$ and the in-plane magnetic field, $B_{||}$, oriented along the [10] direction. For the band $xy$ the $d-$wave component is zero and is not presented here. (f-j) The selected pairing amplitudes for the band $yz$ and the magnetic field orientation [01] and [11]. Results for $J=0.175$~eV ($J'=0.0175$~eV), $\Delta_{SO}=30$~meV and $\Delta_{RSO}=30$~meV. }
\label{fig5}
\end{figure*}

\label{sec3b}
\begin{figure}[!h]
 \centering
 \includegraphics[width=0.3\textwidth]{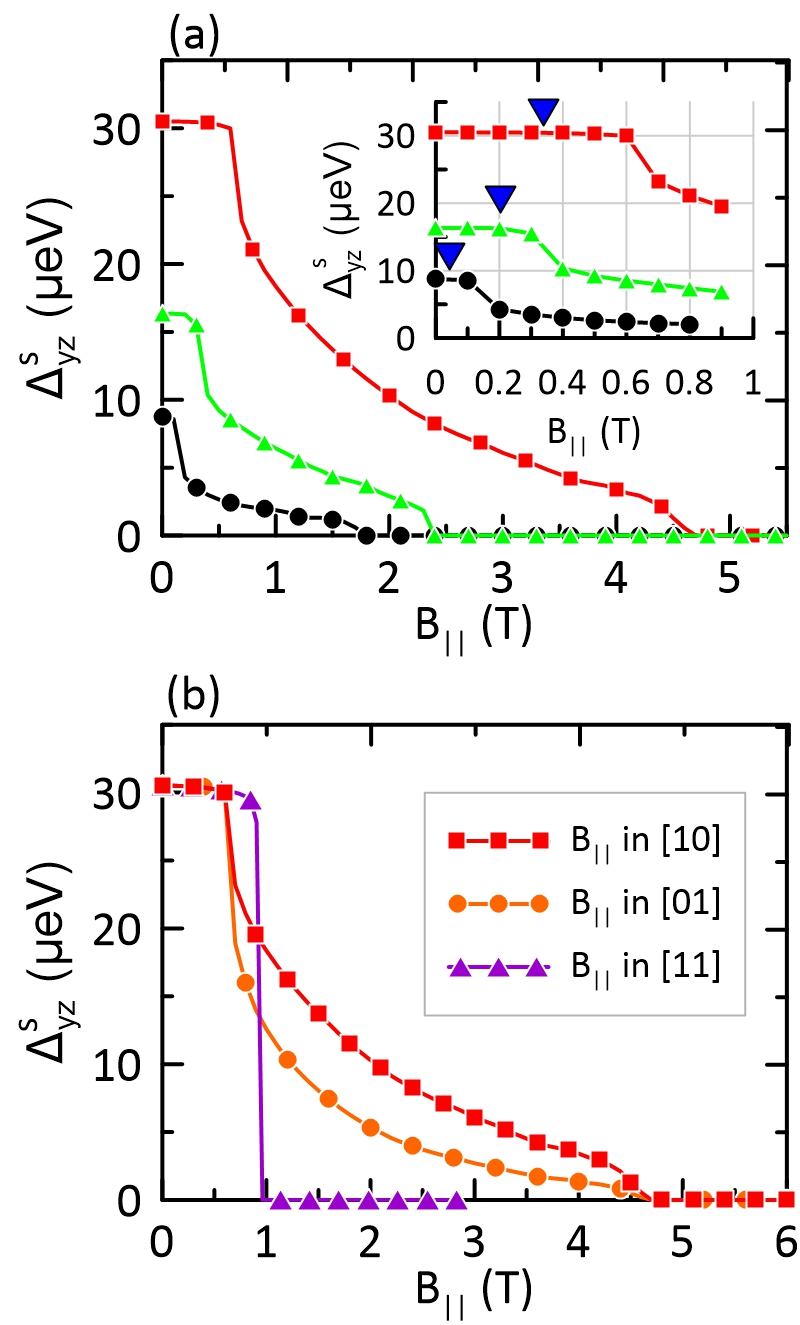}
 \caption{(a) The extended $s-$wave superconducting amplitude for the $yz$ band and three selected values of $n_{tot}$ corresponding to the $B_{c||}$ maximum (red) and two points just below (blue) and much above it (green). The results correspond to the [10] orientation of the field. The inset presents a zoom of the main panel at the low magnetic field regime where the blue triangles denote the paramagnetic limit determined from the Chandrasekhar-Clogston formula. (b) The superconducting amplitudes $\Delta_{yz}^s$ calculated at $n_{tot}$ corresponding to the $B_{c||}$ maxima and three different magnetic field orientations.}
\label{fig6}
\end{figure}
We now analyze the effect of SOC on the in-plane critical magnetic field of the superconducting 2DEG at the LAO/STO interface. 
For a parallel magnetic field, due to the strong confinement of electrons in a few nanometer quantum well close to the interface, the orbital motion and vortices can be neglected, making the coupling of the magnetic field to the spin and the orbital magnetic momentum [cf. $\hat{H}_B$ in Sec.\ref{sec2}] the dominant pair-breaking effect.
In contrast to the case with $B=0$, now the bands $xz$ and $yz$ are no longer equivalent. Instead, their response to the external magnetic field depends on the $\mathbf{B}_{||}$ orientation. Due to the $C_4$ symmetry, the pairing amplitudes for the band $yz$ and the magnetic field directed along $[10]$ are equivalent to the amplitudes for the band $xz$ and the magnetic field along $[01]$. Therefore, for the sake of clarity, in this subsection we mainly concentrate on the band $yz$ and the lower lying band $xy$.
\begin{figure}[!h]
 \centering
 \includegraphics[width=0.5\textwidth]{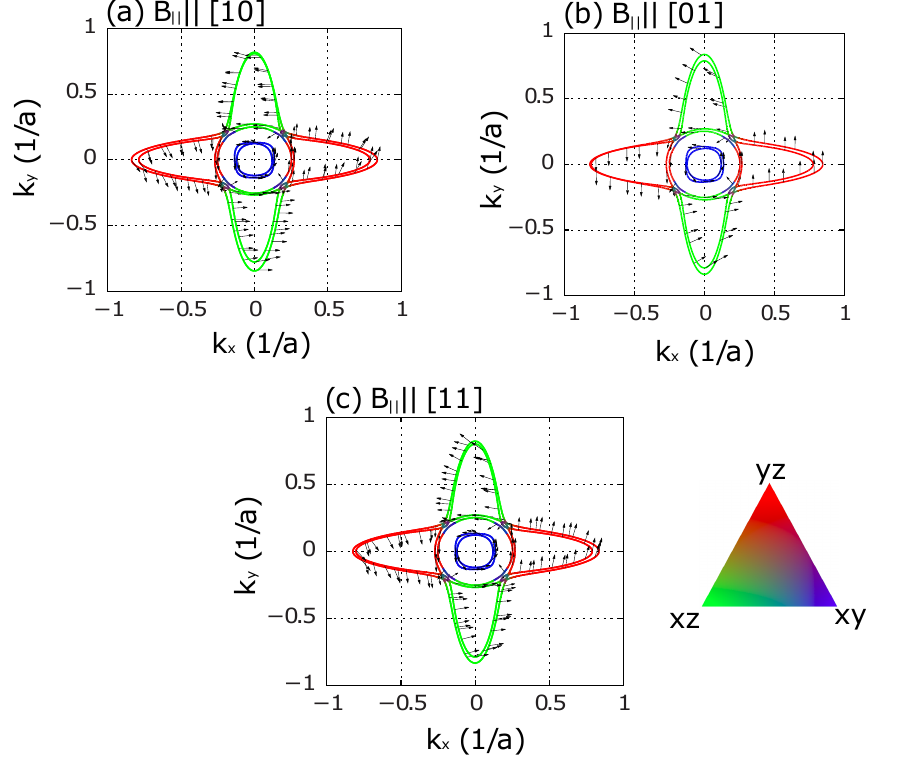}
 \caption{Fermi surfaces calculated for $n_{tot}$ corresponding to the $T_{c}$ maxima and $B_{||}=4$~T directed along (a) [01], (b) [10] and (c) [11]. The contribution of the individual bands are marked in color. Black arrows indicate the spin direction and amplitude. }
\label{fig6_exp}
\end{figure}

For the magnetic field oriented in the [10] direction, the non-monotonic (dome shape) dependence of the critical field at $T=0$,  is shown in Fig.~\ref{fig5}(a-e). It corresponds to the dome of $B_{c||}$ observed experimentally\cite{Rout2017} as a function of the gate voltage. Similarly as before, the dominant pairing amplitude is related to the extended $s-$wave superconducting gap in the bands $xz/yz$ [Fig.~\ref{fig5}(a)]. However, now the additional triplet $p_x-$wave component is induced due to the interplay between $\mathbf{B}$ and SOC. Note that only the triplet component $p_{x,(y)}$ corresponding to the magnetic filed orientation is created, i.e., for $\mathbf{B}_{||}$ along $[10]$ the $p_y-$wave component is zero.

Even though, for different bands the superconducting amplitudes at $B_{||}=0$ have different magnitudes, they are all characterized by a single critical field as the pairings in all bands are linked by the Cooper pair-hopping term. The calculated value of $B_{c||}$, which is slightly above $4$~T at the maximum, quantitatively agrees with the one measured in the (111) LAO/STO interface\cite{Rout2017} that is characterized by the highest SOC strength, close to that assumed in our calculations. Although the critical field for other crystallographic orientations is reported to be slightly smaller\cite{Reyren2007,BenShalom2010}, which we believe is a result of lower SOC strength, our results remain close to the experimental values. Note that as presented in panels (a,f,i), the critical field $B_{c||}$ strongly depends on the magnetic field orientation. For the [01] and [10] direction, $B_{c||}$ at the maximum is almost four times higher compared to the case when the magnetic field is oriented along the [11] direction.

In 2D superconductors, the orbital effects of the in-plane magnetic field are strongly suppressed due to the confinement in the transverse direction. The parallel upper critical field ($B_{c||}$) is determined solely by the Zeeman effect and is given by the Chandrasekhar-Clogston formula\cite{Chandrasekhar1962,Clogston1962}, $B_{c||}=\Delta(B=0) /\sqrt{2}g\mu_B$. However, in the presence of SOC, this upper bound is predicted to be significantly increased\cite{Frigeri2004} due to the coupling of the electron spin with its momentum. When the inversion symmetry is broken, the spin of the electron is no longer free to rotate but is pinned to the effective field of SOC. For the spin-orbit energy much larger than the Zeeman splitting, the paring occurs between the helical states. In this case, the spin flip induced by the magnetic field, which destroys the spin-singlet Cooper pairs, is much less likely due to the spin-momentum locking.
Therefore, we expect that the superconductivity should be more persistent in the presence of the external magnetic field after the inclusion of SOC

In Fig.~\ref{fig6}(a) we present the dominant extended $s-$wave pairing amplitude for the bands $yz$, calculated as a function of the magnetic field oriented in [01]. We select three values of $n_{tot}$, one corresponding to the $B_{c||}$ maximum (red line) and two others just below the maximum, in the characteristic region induced by the atomic SOC (black), and much above the maximum, in the region of the superconducting amplitude drop (green).
Note that the critical field for all of them largely excesses the Chandrasekhar-Clogston limit denoted by blue triangles in the inset.

In Fig.~\ref{fig6}(a) we can distinguish two regimes: the first, where the pairing amplitude is almost constant, and the second, for higher magnetic fields, in the form of a superconducting tail, which is a manifestation of the nontrivial interplay between superconductivity and SOC. Importantly, calculations without SOC (not displayed here) do not exhibit the characteristic tail and the superconductivity is destroyed exactly at the Chandrasekhar-Clogston limit within the first-order transition. 

The superconducting tail and the corresponding critical field both depend on the magnetic field orientation and are different for [01] ([10]) and [11] directions as presented in panel (b).
Interestingly, the tail does not occur when the magnetic field is directed along [11] for which the superconductor-to-normal state transition is of the first order.
This behavior can be explained if we analyze changes of the Fermi surface induced by the magnetic field -- see Fig.~\ref{fig6_exp}. For $\mathbf{B}=0$, the superconducting state is created as a result of pairing between electrons with opposite spins and momenta within the band, $(\mathbf{k},l,\uparrow) \Longleftrightarrow (-\mathbf{k},l,\downarrow)$, where $l=yz,xz,xy$. The superconductivity is not significantly affected by SOC which does not break the time reversal symmetry, i.e. $E(\mathbf{k},l,\uparrow)=E(-\mathbf{k},l,\downarrow)$.
When the magnetic field is applied along [10], the spin and orbital Zeeman effect leads to the wave vector mismatch between electrons with opposite spins in the band $xz$, one of the main (together with $yz$) bands responsible for the superconducting state -- see Fig.~\ref{fig6_exp}(a). This effect leads to depairing of the Cooper pairs in the band $xz$. Note, however, that the wave vector mismatch is not induced in the band $yz$ in which the pairing of electrons with opposite spin and momenta is still possible. This pairing in a part of the Fermi surface results in the superconducting tail presented in Fig.~\ref{fig6}(b). 
The analogous effect occurs when the magnetic field is applied along [01]. Then, the wave vector mismatch appears only in the band $yz$, with the conditions for the pairing being unaffected for the band $xz$ -- see Fig.~\ref{fig6_exp}(b). Note that the pairing is destroyed in both the bands $yz$ and $xz$ by the wave vector mismatch when the magnetic field is directed along [11]. In this case, the superconductor-to-normal phase transition does not exhibit a tail but has a step-like form characteristic for the first-order transition. 
\begin{figure}[t]
 \centering
 \includegraphics[width=0.5\textwidth]{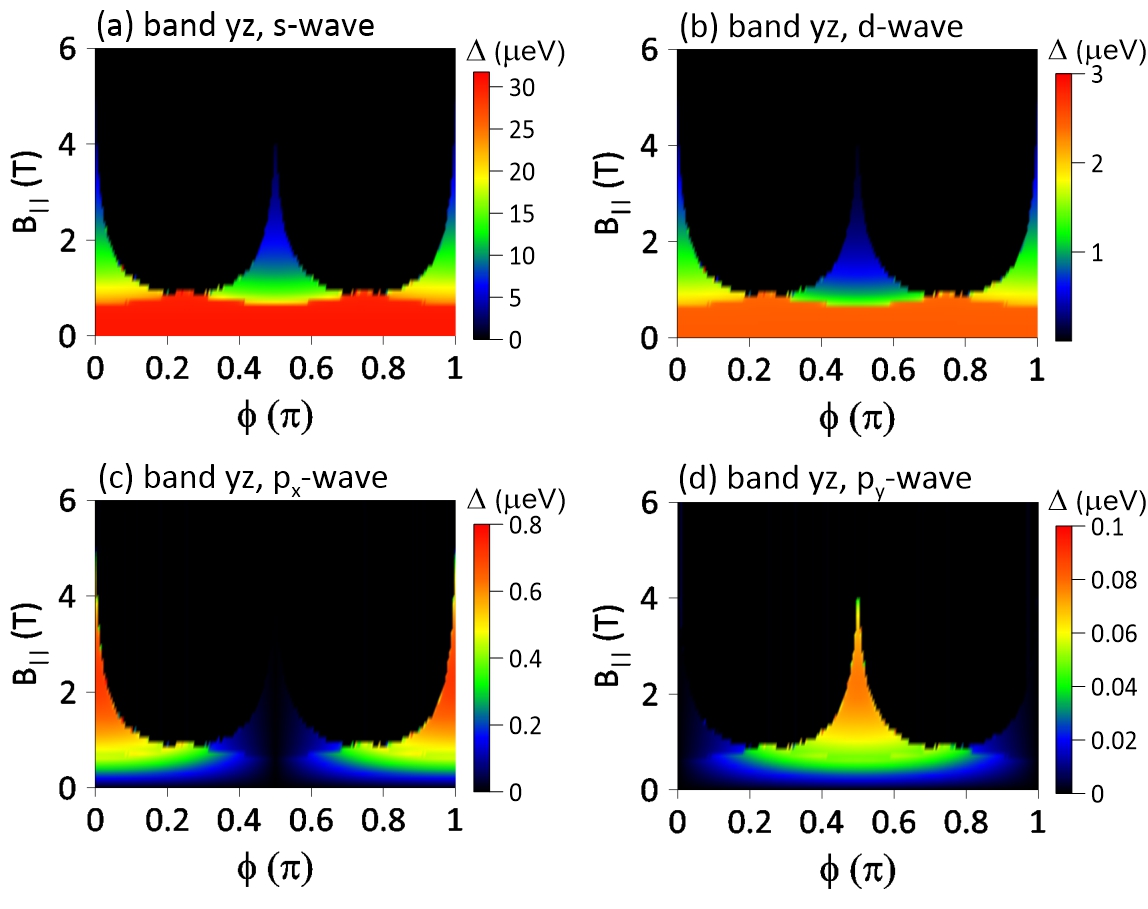}
 \caption{The pairing amplitudes (a) $\Delta_{yz}^s$, (b) $\Delta_{yz}^d$, (c) $\Delta_{yz}^{p_x}$ and (d) $\Delta_{yz}^{p_y}$ as a function of the magnetic field intensity $B_{||}$ and direction, defined by the angle $\phi$ measured relative to the [10] axis.}
\label{fig7}
\end{figure} 

Finally, one should also note that the Fermi surface in LAO/STO is characterized by the $C_4$ symmetry at $\mathbf{B}=0$. This, in turn, results in the four-fold anisotropy of the critical field, demonstrated in Fig.~\ref{fig7} where the dominant pairing amplitudes for the band $yz$ are presented as a function of the magnetic field intensity $B_{||}$ and direction, defined by the angle $\phi$ measured in relation to the [10] axis. Note that indeed the enhancement of the critical field is characteristic for the high symmetry directions [10] and [01] while $B_{c||}$ gradually decreases when the magnetic field orientation approaches the direction [11]. Interestingly, the triplet $p_{x(y)}$ components disappear in the magnetic field direction [01] and [10], respectively.
\begin{figure}[t]
 \centering
 \includegraphics[width=0.5\textwidth]{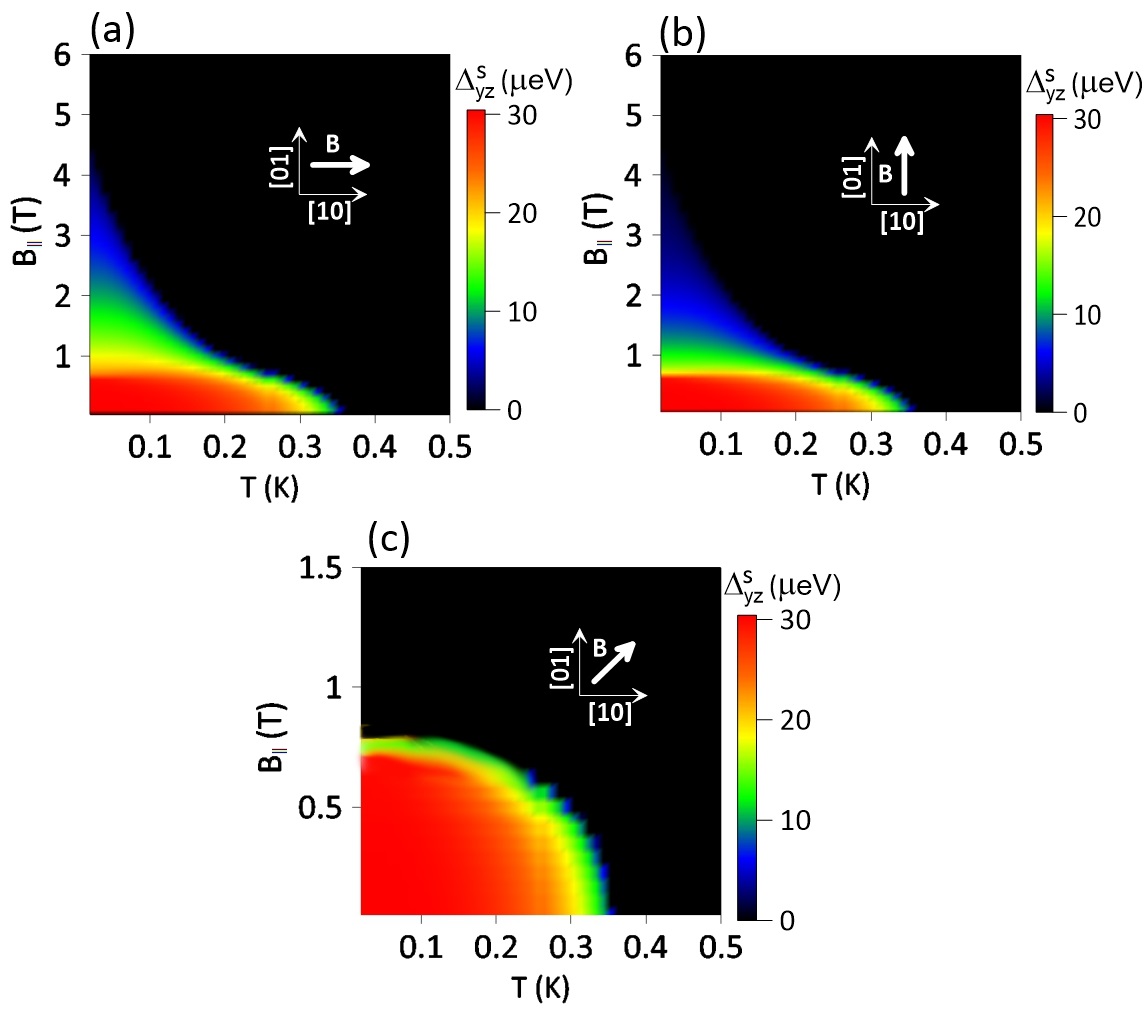}
 \caption{Temperature dependence of the pairing amplitude $\Delta_{xz/yz}^s$ calculated at $n_{tot}$ corresponding to the $T_{c}$ maxima for different magnetic field orientations.}
\label{fig8}
\end{figure} 

For the sake of completeness, we also calculate the temperature dependence of the paring amplitudes at $n_{tot}$ corresponding to the $T_{c}$ maxima (Fig.~\ref{fig8}).
As can be seen, the enhancement of the critical field and the corresponding $B_{c||}$ anisotropy is prominent at the low temperature regime. For higher $T$ the superconducting tail gradually disappears significantly weakening the critical field and reducing the $B_{c||}$ anisotropy.

\section{Conclusions}
\label{sec4}
Assuming the real space nearest neighbour pairing, we have studied the impact of the SOC on the phase diagram of the LAO/STO interface within the mean field Hartree-Fock approximation. 
The atomic and Rashba SOC have been considered separately and, as shown, both are detrimental to the superconducting state.

According to our analysis, due to SOC the system undergoes two Lifshitz transitions within the low-carrier concentration regime. Superconductivity is created slightly above the first one when the Fermi energy passes the anticrossing between the bands $xy$ and $yz$. Then, a rapid enhancement of superconductivity occurs at the second Lifshitz transition, with the maximal $T_c$ being reached slightly above it. Therefore, the distance between the lower critical concentrations for the appearance of the superconducting state and the optimal concentration, for which the maximal $T_c$ appears, is largely influenced by the atomic SOC energy. The Rashba interaction, on the other hand, only weakly suppresses the superconductivity and for its large value, can lead to the characteristic narrow peak in the superconducting amplitudes resulting from the van Hove singularity. It would be interesting to determine which of the two Lifshitz transitions are seen in experiment. Possibly different positions of LT reported experimentally\cite{joshua2012universal,Biscaras} may actually correspond to LT1 and LT2 which appear in our analysis.

In the presence of the magnetic field $B_{c,||}$ exhibits the characteristic dome with its critical values close to the ones measured experimentally~\cite{Rout2017}. 
The $C_4$ symmetry of the Fermi surface results in a four-fold anisotropy of the in-plane critical field with respect to the magnetic field orientation. For high symmetry directions, the estimated value of $B_{c||}$ exceeds the paramagnetic limit several times. This effect, reported also in experiments\cite{Rout2017}, is a manifestation of the interplay between superconductivity and the SOC and occurs for [01] and [10] magnetic field orientations. The characteristic superconducting tail in this case is explained as resulting from depairing of the Cooper pairs in a fraction of the Fermi surface where the wave vector mismatch appears. 

Finally, we should note that when the in-plane magnetic field is applied, the center of the band (Fermi surface) is shifted away from the $\Gamma$ point [\ref{fig6_exp}]. In this case, it could be energetically favorable to create the nonzero total momentum of Cooper pairs with a pairing between electrons in the states $| \mathbf{k} \uparrow \rangle$ and $| -\mathbf{k}+\mathbf{q} \downarrow \rangle$. This would lead to a superconducting condensate with a spatial $e^{2i\mathbf{q}\cdot \mathbf{r}}$ modulation of the gap, similarly as in the Fulde-Ferrell-Larkin-Ovchnnikov state\cite{Fulde1964,Larkin1964,}, and entails the nontrivial coupling between the magnetic field and the supercurrent. 
Note, however, that the nonzero total momentum pairing is not obeyed by the Anderson theorem and so it is easily destroyed by any imperfections of the sample. This makes the so-called helical state extremely unstable\cite{Wang2018} and for this reason we neglect it in the present paper.

As a final remark, we should mention that the extended $s$-wave symmetry is in all considered cases the dominant component of the gap at LAO/STO interface. It reproduces the experimental phase diagram with a good accuracy. Nevertheless, for the case with both SOC coupling and the magnetic field present, a small $p_x/p_y$-wave pairing appears. Although the physical mechanism behind the superconductivity in LAO/STO is still under debate, recently two theoretical proposals have been reported, which are consistent with our hypothesis on the extended $s$-wave symmetry of the gap. According to the first scenario\cite{Pekker2020} the electron pairing is mediated by the ferroelastic domain walls which couple to the electron density leading to the alternatively occurring electron-rich and electron-poor regions. The low-energy excitation at the LAO/STO interface results in superconductivity around the edges of electron-rich regions with real-space intersite pairing mechanism. This mechanism stabilizes the extended $s$-wave superconducting state similarly as in our case. The second scenario is based on the fluctuations of momentum-based multipoles as analyzed in Ref.\cite{Sumita2020}. Within such a concept, the interaction vertex under the crystal symmetry corresponding to STO reveals an extended $s$-wave symmetry of the electron pairing.

\section{Acknowledgement}
The work was supported by National Science Centre (NCN) according to the decision number 2017/26/D/ST3/00109 and in part by PL-Grid Infrastructure.

%

\end{document}